\documentclass[review]{elsarticle}
\usepackage{tabularx}

\usepackage{color, colortbl}

\usepackage{lineno,hyperref}

\journal{Journal of \LaTeX\ Templates}

\definecolor{Gray}{gray}{0.9}
\begin{document}

\begin{frontmatter}
	\title{Magnetic couplings and magnetocaloric effect in the GdTX (T=Sc, Ti, Co, Fe; X=Si,Ge) compounds}

\author{Daniel J. Garc\'{\i}a$^{1,2}$,Ver\'onica Vildosola $^{1,3}$, Pablo S. Cornaglia$^{1,2}$}

\address{$^1$ Consejo Nacional de Investigaciones Cient\'{\i}ficas y T\'ecnicas (CONICET), Argentina}
\address{$^2$ Centro At{\'o}mico Bariloche and Instituto Balseiro, CNEA, 8400 Bariloche, Argentina}
\address{$^3$ Centro At{\'o}mico Constituyentes, CNEA, Buenos Aires, Argentina}

\begin{abstract}

We compute the magnetocaloric effect (MCE) in the  GdTX (T=Sc, Ti, Co, Fe; X=Si, Ge) compounds as a function of the temperature and the external magnetic field.
To this end we use a density functional theory approach to calculate the exchange-coupling interactions between Gd$^{3+}$ ions on each compound.
We consider a simplified magnetic Hamiltonian and analyze the dependence of the exchange couplings on the transition metal T, the p-block element X, and the crystal structure (CeFeSi-type or CeScSi-type). 
The most significant effects are observed for the replacements Ti $\to$ Sc or Fe $\to$ Co which have an associated change in the parity of the electron number in the 3d level.
These replacements lead to an antiferromagnetic contribution to the magnetic couplings that reduces the Curie temperature and can even lead to an antiferromagnetic ground state. 
We solve the magnetic models through mean field and Monte Carlo calculations and find large variations among compounds in the magnetic transition temperature
and in the magnetocaloric effect, in agreement with the available experimental data. 
The magnetocaloric effect shows a universal behavior as a function of temperature and magnetic field in the ferromagnetic compounds after a scaling of the relevant energy scales by the Curie temperature $T_C$.
\end{abstract}

\begin{keyword}
Magnetism\sep Rare earth\sep DFT\sep Magnetocaloric effect 
\end{keyword}

\end{frontmatter}


\section{Introduction}
Gadolinium based compounds, in particular the GdTX family \cite{gupta2015review}, have been the subject numerous theoretical\cite{cremades2012theoretical,Liu2013gdscsi,liu2010gdfesi,talakesh2017density} and experimental\cite{PhysRevB.99.134429,pecharsky2003giant,pecharsky1999magnetocaloric,du2008large,luo2018exploring,guillou2017crystal} studies because of their potential use in refrigeration at room temperature using the magnetocaloric effect. The strong magnetocaloric properties of Gd systems are due to the large spins in the 4f Gd$^{3+}$ ions and the exchange couplings between them that lead to magnetic phase transitions. A giant magnetocaloric effect is observed in Gd$_5$(Si$_2$Ge$_2$) at room temperature, associated with the presence of a first order ferromagnetic (I) $\leftrightarrow$ ferromagnetic (II) transition at $T\simeq 276K$ \cite{pecharsky1997giant}. Pure gadolinium, which is also a strong magnetocaloric material at room temperature, presents a second order Curie transition at $T_C=294$K to a ferromagnetic ground state.

The magnetocaloric effect (MCE) is generally quantified by the entropy change $\Delta S_M=S(H)-S(0)$, when an external magnetic effect $H$ is applied. To obtain a large MCE a high sensitivity of the material to magnetic field changes is required. This is the case, e.g., close to a paramagnetic-ferromagnetic transition temperature where the MCE acquires its maximum values. For magnetic cooling applications it is important to maximize the MCE at the operation temperatures. A route to attain this goal is to control the magnetic transition temperature through the magnetic couplings which are determined by the conduction band structure and its occupancy. 

The ternary RTX compounds (where R is a rare earth), T is a transition metal and X is a p-block element such as Si, Ge, Sb) present several examples of large MCE compounds. GdFeSi is a ferromagnet below $T_C=118K$ where the MCE attains its maximum value $\Delta S_M=-22.3J/KgK =-0.42R$ ($\Delta H=9 \textrm{Tesla}$), GdScSi and  GdScGe are also ferromagnets with $T_C=318$K and $\Delta S_M=-2.5J/KgK=-0.047R$ and $T_C=320$K and $\Delta S_M=-3.3J/KgK=-0.62R$ for $\Delta H=2 \textrm{Tesla}$, respectively. GdCoSi is however an antiferromagnet with $T_N=220$K and a low MCE \cite{skorek2001electronic,couillaud2011magnetocaloric,napoletano2000magnetic}.

\begin{figure}[t]
    \begin{center}
        \includegraphics[width=0.6\textwidth]{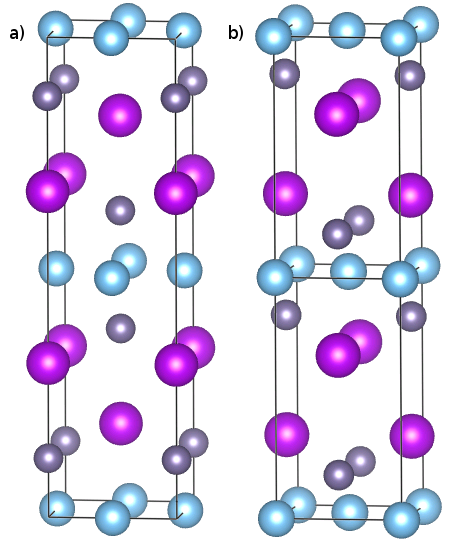}
    \end{center}
	\caption{(Color online) Crystal structures: a) CeScSi-type b) CeFeSi-type. The rare earth atoms are represented by the larger spheres, the p-block atoms by the smallest spheres and the transition metal atoms by the middle size spheres.}
    \label{fig:unitcell}
\end{figure}

Among the variety of RTX crystal structures, we will focus on the tetragonal CeFeSi-type (space group P4/nmm) and  CeScSi-type (space group I4/mmm) R-X-T$_2$-X-R structures (see Fig. \ref{fig:unitcell}) and analyze the role of T and X in the magnetic properties for the R=Gd case. As we show below, the results of this analysis will prove helpful interpreting the experimental results for other rare earths.

\section{Magnetic ground state and coupling constants}
Total-energy calculations of the GdTX compounds, based on density functional theory (DFT), indicate a ground state with magnetic moments localized at the Gd$^{3+}$ ions and allowed us to estimate the strength of the Gd-Gd magnetic interactions. We solved the resulting magnetic model to obtain the magnetic contribution to the specific heat, the magnetocaloric effect, and the N\'eel and Curie transition temperatures.

\subsection{Technical details of the DFT calculations} 
The total-energy calculations were performed using the generalized gradient approximation (GGA) of Perdew, Burke and Ernzerhof for the exchange and correlation functional as implemented in the Wien2K code \cite{wien2k,Perdew1996}.
A local Coulomb repulsion was included in the Gd $4f$ shell and treated using GGA+U which is a reasonable approximation for these highly localized states. 
Due to the localized character of the $4f$ electrons, the fully localized limit was used for the double counting correction \cite{Anisimov1993}. 
We described the local Coulomb and exchange interactions with a single effective local repulsion $U_{eff} = U - J_H = 6 eV$,  which has been successfully used in bulk Gd \cite{Yin2006,Petersen2006}.
The APW+local orbitals method of the \textsc{WIEN2K} code was used for the basis function \cite{wien2k}. We used 1200 k-points in the Brioullin zone for the full optimization of the crystal structures, and 200 k-points for the $2\times2\times2$ supercell total-energy calculations of the different magnetic configurations. The lowest energy configuration is identified as the magnetic ground state, which in all the analyzed cases corresponds to the type of order experimentally observed \cite{welter1994proprietes,klosek2002magnetic,WELTER199249,welter1994magnetic,nikitin1998magnetic,gaudin2011,nikitin2002magnetic}. 
The magnetic moments are localized on the Gd 4f orbitals and no significant magnetic moment is obtained in the transition metal (for T=Fe, Co, Sc, Ti, except in the ferromagnetic state where a small spin polarization is observed), in agreement with neutron diffraction experiments \cite{WELTER199249}. 

\begin{table}
	\centering
    \begin{tabularx}{\columnwidth}{@{}l *8{>{\centering\arraybackslash}X}@{}}
		\hline
		\hline
		& {\underline{GdFeSi}} &\underline{GdCoSi}&\underline{GdTiSi}&\underline{GdTiGe} &GdTiGe&GdScGe&GdScSi \\
		\hline
		{\bf FM} &\cellcolor{Gray} 0   & 198& 305&424& \cellcolor{Gray}   0 & \cellcolor{Gray}0    &\cellcolor{Gray}0\\
		AF1& 43	 & 891& 1608&1459& 1967 & 1551 &1537\\
		{\bf AF2}& 220& \cellcolor{Gray}  0 & \cellcolor{Gray}  0& \cellcolor{Gray}0& 1013 & 377  &370\\
		AF3& --	 & 711& 1339&1442 &1941 & 1220 &1217\\
		AF4& 466 & 788& 1609&1552 &2197 & 1564 &1546\\
		AF5& 244 & 661& 1183&1281 &2001 & 1223 &1200\\
		AF6& 243 & 660& 1183&558 &1942 & 1220 &1217\\
		AF7& 302 & 704& 1572&1476 &2071 & 1462 &1431\\
		\hline
		\hline
	\end{tabularx}
	\caption{Relative energy $\Delta E$ (in K) with respect to the ground state for the magnetic configurations of Fig. \ref{fig:configurations} for a DFT cell with 16 Gd$^{3+}$ ions. The AF3 configuration is unstable for GdFeSi. The underlined compounds correspond to the CeFeSi-type structure, and the rest to the CeScSi-type structure.}
	\label{tab:totnrg}
\end{table}
\subsection{Magnetic structure of the ground state and coupling constants}
\begin{figure}[h]
  \centering
    \includegraphics[width=\textwidth]{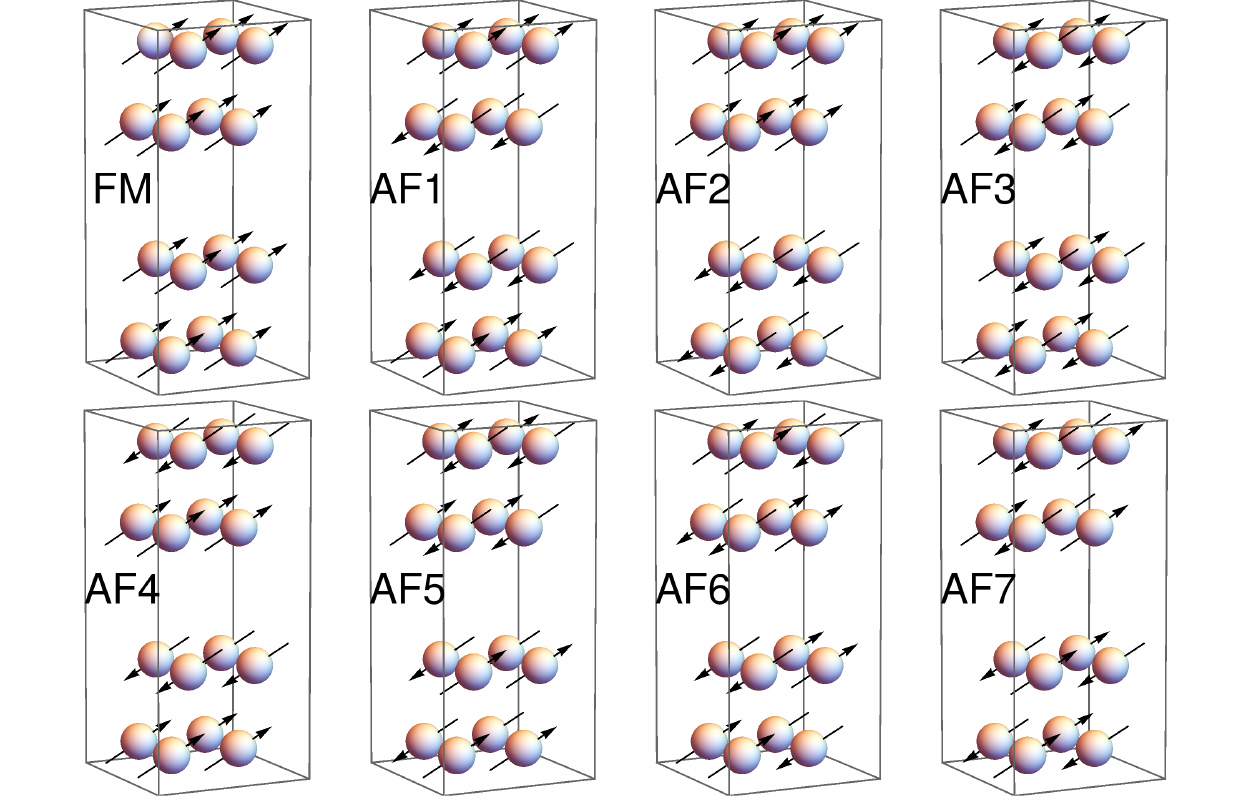}
    \caption{(Color online) Magnetic configurations proposed for the CeFeSi-type structures to determine the ground state and obtain the exchange coupling parameters. The spheres correspond to Gd ions and the orientation of the Gd$^{3+}$ 4f magnetic moments is indicated by black arrows. A analogous set of magnetic configurations was used for the CeScSi-type structures. }
  \label{fig:configurations}
\end{figure}

We explored different static configurations for the magnetic moments which are presented in Fig. \ref{fig:configurations} for the CeFeSi-type structures. The magnetic configurations used for the CeScSi-type structures are completely analogous, with the same relative orientation of the magnetic moments inside each Gd layer and between the layers.  
The relevant quantity to study the magnetic interactions is the energy difference between magnetic configurations. 
The energy differences between a given magnetic moment configuration and the lowest energy one are presented, for each compound, in Table \ref{tab:totnrg}. 
A ferromagnet for GdFeSi, GdTiGe (I4/mmm), GdScGe and GdScSi, and an A-type antiferromagnet for GdCoSi, GdTiSi, and GdTiGe (P4/nmm) with the magnetic moments of the Gd on each bilayer aligned ferromagnetically, and each bilayer aligned antiferromagnetically with its nearest neighbouring bilayers. 

In these metallic compounds, the dominant Gd-Gd interactions are due to a Ruderman-Kittel-Kasuya-Yosida (RKKY) coupling between the Gd's magnetic moments through an exchange interaction with the conduction electrons, which decay in 3D systems as an inverse third power of the inter Gd distance~\cite{PhysRev.106.893,kasuya1956,PhysRev.96.99}. 
DFT calculations of the RKKY couplings in GdFeSi in Ref. \cite{Liu2010first,liu2010gdfesi} suggests an even faster decay with increasing inter Gd distance. 
As it is customary we considered a finite set of exchange interactions. The seven magnetic configurations considered allow us to calculate up to 6 exchange couplings. We analyzed several choices for the non-zero exchange couplings and found, in all cases, that there was no lower energy magnetic configuration, compatible with the obtained couplings, than the already considered.

\begin{figure}[h]
  \centering
    \includegraphics[width=0.7\textwidth]{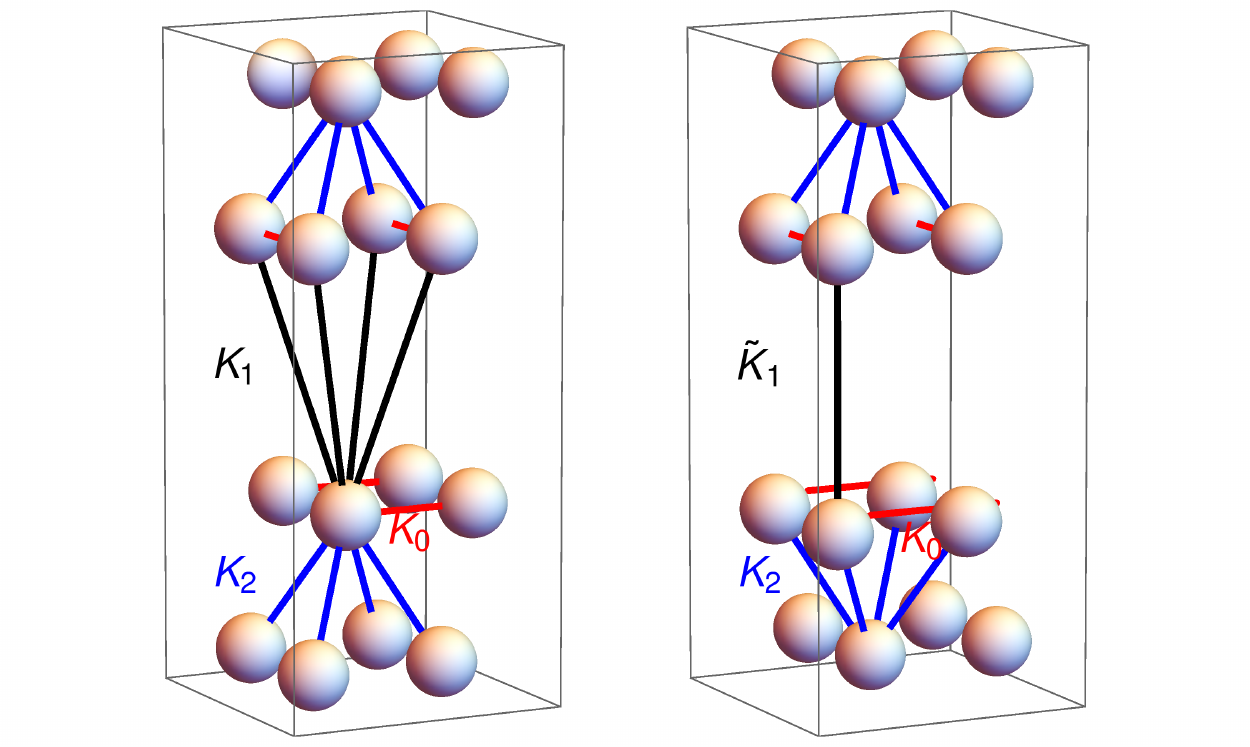}
    \caption{(Color online)  Magnetic couplings considered in the simplified model. The lines connect pairs of Gd atoms that are magnetically coupled (to avoid overloading the plot, not all coulings are drawn, but can be inferred from symmetry considerations) through the exchange coupling parameters $K_0$, $K_1$ (only for the CeFeSi-type structure), $\tilde{K}_1$ (only for the CeScSi-type structure), and $K_2$, as indicated in the figure. }
  \label{fig:couplings}
\end{figure}

We found that an accurate description of the system is obtained using a simplified model for the magnetic interaction between Gd$^{3+}$ magnetic moments, with three coupling constants (see Fig. \ref{fig:couplings}) \ref{WELTER199954,klosek2002magnetic}. An exchange coupling $K_0$ between nearest neighbour Gd atoms on each Gd layer (which is a square lattice), a coupling $K_2$ between nearest neighbours in different layers of the bilayer, and a nearest neighbour coupling between Gd in different bilayers. For the latter coupling there are two possibilities depending on the lattice type: $K_1$ associated with 4 neighbours in the CeFeSi-type structure, and $\tilde{K}_1$ associated with a single neighbour in the CeScSi-type structure (see Fig. \ref{fig:couplings}).
We found that including up to three additional magnetic couplings in the model only lead to minor quantitative differences in the calculated properties.

In the absence of an applied magnetic field, the contribution per Gd atom to the total energy due to the magnetic interactions, described in Fig. (\ref{fig:couplings}), for the different configurations of Fig. \ref{fig:configurations} is given for the two structures considered in Table \ref{tab:energiesH}:

\begin{table}
    \centering
    \begin{tabularx}{\columnwidth}{@{}l *3{>{\centering\arraybackslash}X}@{}}
\hline
	& CeFeSi-type & CeScSi-type\\
\hline
\hline
	$E^m_{FM}/J^2 $  & $-4 ({K}_0 + {K}_1 + {K}_2)$ &$- 4 {K}_0 - \tilde{K}_1 -4 {K}_2$  \\
	$E^m_{AF1} /J^2$ & $-4 ({K}_0 + {K}_1 - {K}_2)$&$-4 {K}_0 - \tilde{K}_1 +4 {K}_2$ \\
	$E^m_{AF2} /J^2$ & $-4 ({K}_0 - {K}_1 + {K}_2)$&$-4 {K}_0 + \tilde{K}_1 -4 {K}_2$ \\
	$E^m_{AF3} /J^2$ & $0$ & $-\tilde{K}_1$ \\ 
	$E^m_{AF4} /J^2$ & $-4 ({K}_0 - {K}_1 - {K}_2)$&$-4 {K}_0 + \tilde{K}_1 + 4 {K}_2$ \\
	$E^m_{AF5} /J^2$ & $0$ & $ \tilde{K}_1$ \\
	$E^m_{AF6} /J^2$ & $0$ & $-\tilde{K}_1$ \\
	$E^m_{AF7} /J^2$ & $4 {K}_0$&$4 {K}_0 +\tilde{K}_1$\\
\hline
    \end{tabularx}
    \caption{Magnetic energy as a function of the magnetic-exchange couplings for the different magnetic configurations considered in the CeFeSi-type and the CeScSi-type structures. $J=7/2$ is the angular momentum of the Gd$^{3+}$ ion $4f$ electrons.}
    \label{tab:energiesH}
\end{table}

The energy differences between magnetic configurations calculated from first principles can be combined with Table \ref{tab:energiesH} to obtain the coupling parameters through a least squares analysis.
The obtained couplings for the different compounds are presented in Table \ref{tab:exchcoup}. For all the studied compounds, the intrabilayer couplings $K_0$ and $K_2$ are positive which indicates that in all cases the Gd magnetic moments in a given bilayer order ferromagnetically. The interbilayer couplings can be positive as in GdFeSi, GdTiGe (I4/mmm), GdScGe, and GdScSi leading to a ferromagnetic ground state or negative as in GdCoSi, GdTiSi, and GdTiGe (P4/nmm) which results in a A-type antiferromagnet in which neighbouring bilayers have their magnetic moments pointing in opposite directions.     
The replacement Si$\rightarrow$Ge does not lead to a significant change in the exchange couplings of GdTiSi and GdScSi which is consistent with the very weak change in the transition temperatures observed in these compounds upon Si$\to$Ge replacement. 
The replacements Fe$\to$Co in GdFeSi and Ti$\to$Sc in GdTiGe produce, however, a change in the sign of the interbilayer coupling $K_1$ and a strong reduction of $\tilde{K}_1$, respectively. These replacements have in common a change in the electron number provided by the transition metal atom, which changes the conduction band occupancy and the RKKY couplings. A double exchange coupling between the Gd magnetic moments through the 3d level of the transition metal naturally gives a change in the sign of the resulting coupling when the 3d level occupancy changes by one electron \cite{oropesa2019minimal}, which may explain the observed behavior of the interbilayer couplings when the transition metal is replaced.

The compound GdTiGe is stable in both the CeFeSi-type (P4/nmm) and CeScSi-type (I4/mmm) structures, but its magnetic behavior depends strongly on the type of structure. GdTiGe (P4/nmm) is an A-type antiferromagnet while GdTiGe (I4/mmm) is a ferromagnet. Although the interbilayer couplings are expected to change because of the different topology, the intrabilayer couplings also change and are roughly twice as large in the CeScSi-type structure.

\begin{table}
    \centering
    \begin{tabularx}{\columnwidth}{@{}l *8{>{\centering\arraybackslash}X}@{}}
        \hline
        \hline
	&\underline{GdFeSi}& \underline{GdCoSi} &\underline{GdTiSi}& \underline{GdTiGe}& GdTiGe&GdScGe&GdScSi\\
        \hline
	${K}_0$& $1.6\,$&$4.4\,$&$10.3\,$&$8$ & $17.0\,$&$10.6$&$10.4$ \\
	${\mathbf{{K}_1}}$  &  $\mathbf{4.5}\,$&$\mathbf{-2.4}\,$&$\mathbf{-3.2}\,$&$\mathbf{-4.1}$& $-\,$&$-$&$-$\\
	${\mathbf{\tilde{K}_1}}$ &  $-\,$&$-\,$&$-\,$&$-\,$& $\mathbf{26.5}\,$&$\mathbf{9.2}\,$&$\mathbf{8.4}$\\
	${K}_2$&  $1.7\,$&$11.5\,$&$21.9$&$18.7$ & $30.9\,$&$23.7$&$23.5$ \\
        \hline
        \hline
	$T_{c}^{MF}$  &\cellcolor{Gray}$163$  & $386$ &$745$&$646$ &$1145$ \cellcolor{Gray} & $766$ \cellcolor{Gray} &\cellcolor{Gray} $756$ \\
	$T_{c}^{MC}$  &\cellcolor{Gray} $117.5$ & $284.5$ &$554.5$&$477$ &\cellcolor{Gray}$637$   & \cellcolor{Gray}$404$  &\cellcolor{Gray} $395$ \\
	$T_{c}^{QMC}$ &\cellcolor{Gray}$154$  & $363$ &$710$&$620$ &$840$ \cellcolor{Gray}  & \cellcolor{Gray}$555$  &\cellcolor{Gray} $530$\\
	$T_{c}^{exp}$ &\cellcolor{Gray} $118^a$ & $175^b$& $400^c$&$412^d$&$376^e$ \cellcolor{Gray}  & $320^f$\cellcolor{Gray}  &$318^f$\cellcolor{Gray} \\
        \hline
        \hline
    \end{tabularx}
    \caption{Calculated exchange couplings (in K) and the associated mean-field Curie-Weiss $\theta$ and N\'eel $T_N^{MF}$ temperatures. 
    Boldface indicates interplane couplings. Shaded cells correspond to Curie temperatures.  The experimental N\'eel $T_N^{exp}$ and Curie-Weiss $\theta^{exp}$ temperatures are presented as a reference. The superscripts indicate the references from which the experimental values were extracted: $a=$\cite{napoletano2000magnetic}, $b=$\cite{welter1994magnetic}, $c$=\cite{klosek2002magnetic}, $d$= \cite{nikitin1998magnetic}, $e=$\cite{gaudin2011}, $f=$\cite{nikitin2002magnetic}.}
    \label{tab:exchcoup}
\end{table}

\section{Magnetocaloric properties}
\begin{figure}[t]
    \begin{center}
      \includegraphics[width=0.6\textwidth]{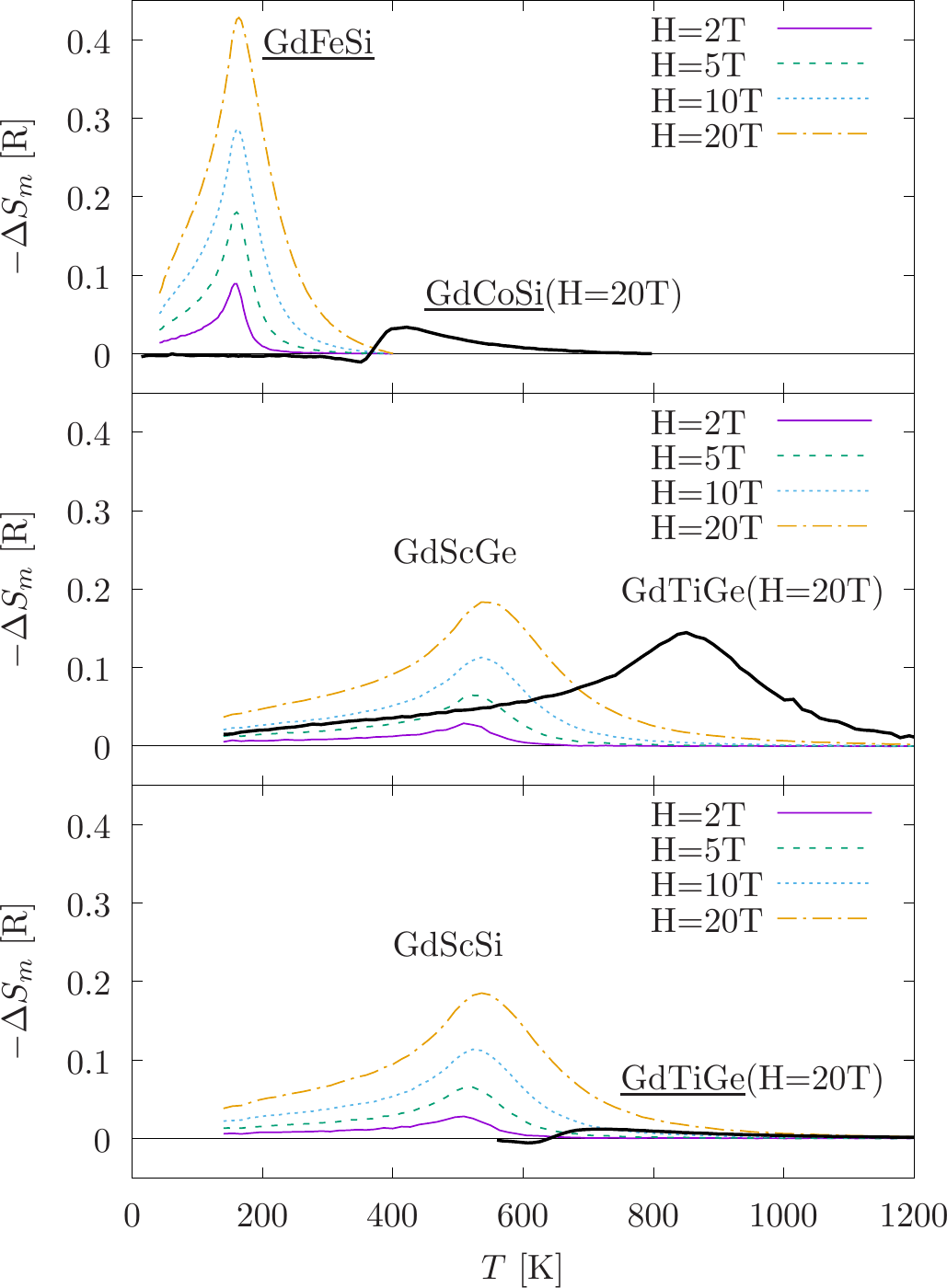}
    \end{center}
    \caption{(Color online) Magnetocaloric effect as a function of the temperature for different compounds and external magnetic fields. The underlined compound formulas correspond to the P4/nmm symmetry. }
    \label{fig:mce}
\end{figure}
We performed a mean-fied (MF) analysis and classical (CMC) and quantum Monte Carlo (QMC) calculations using the obtained magnetic model couplings for the compounds of Table \ref{tab:exchcoup}. We used the ALPS library (see Refs. \cite{ALBUQUERQUE20071187,Bauer_2011}) for the numerical calculations with system sizes of up to $8\times 8\times 8$ magnetic moments. 
In Fig. \ref{fig:mce} we present $-\Delta S_m=S_m(0.1 \textrm{Tesla})-S_m(H)$ calculated numerically using Quantum Monte Carlo, as a function of the temperature for four ferromagnetic and two antiferromagnetic compounds. The ferromagnetic compounds show a peak in $-\Delta S_m$, for temperatures near the Curie temperature, whose height increases monotonically with increasing magnetic field. The maximum $-\Delta S_m$ increases with decreasing $T_C$ while the width of the peak follows an opposite trend.
The antiferromagnetic compounds show a much lower overall intensity of the MCE and a change in the sign of $-\Delta S_m$ near the N\'eel transition.

In the mean-field approximation the scaling $H\to g \mu_B H/(k_B T_C)$ (where $g=2$ is the giromagnetic factor) and $T\to T/T_C$ results in a universal curve for $\Delta S_m$ for the ferromagnetic compounds\cite{bonilla2010universal,franco2008universal,smith2014scaling,biswas2013universality}. In the AFM compounds the maximum value of $-\Delta S_m$ depends strongly on the value of the antiferromagnetic bilayer coupling ($K_1$ or $\tilde{K}_1$ depending on the crystal structure), since the Zeeman energy needs to be large enough to overcome it in order to be able to generate a sizable magnetization and the associated entropy change.
\begin{figure}[t]
    \begin{center}
      \includegraphics[width=0.6\textwidth]{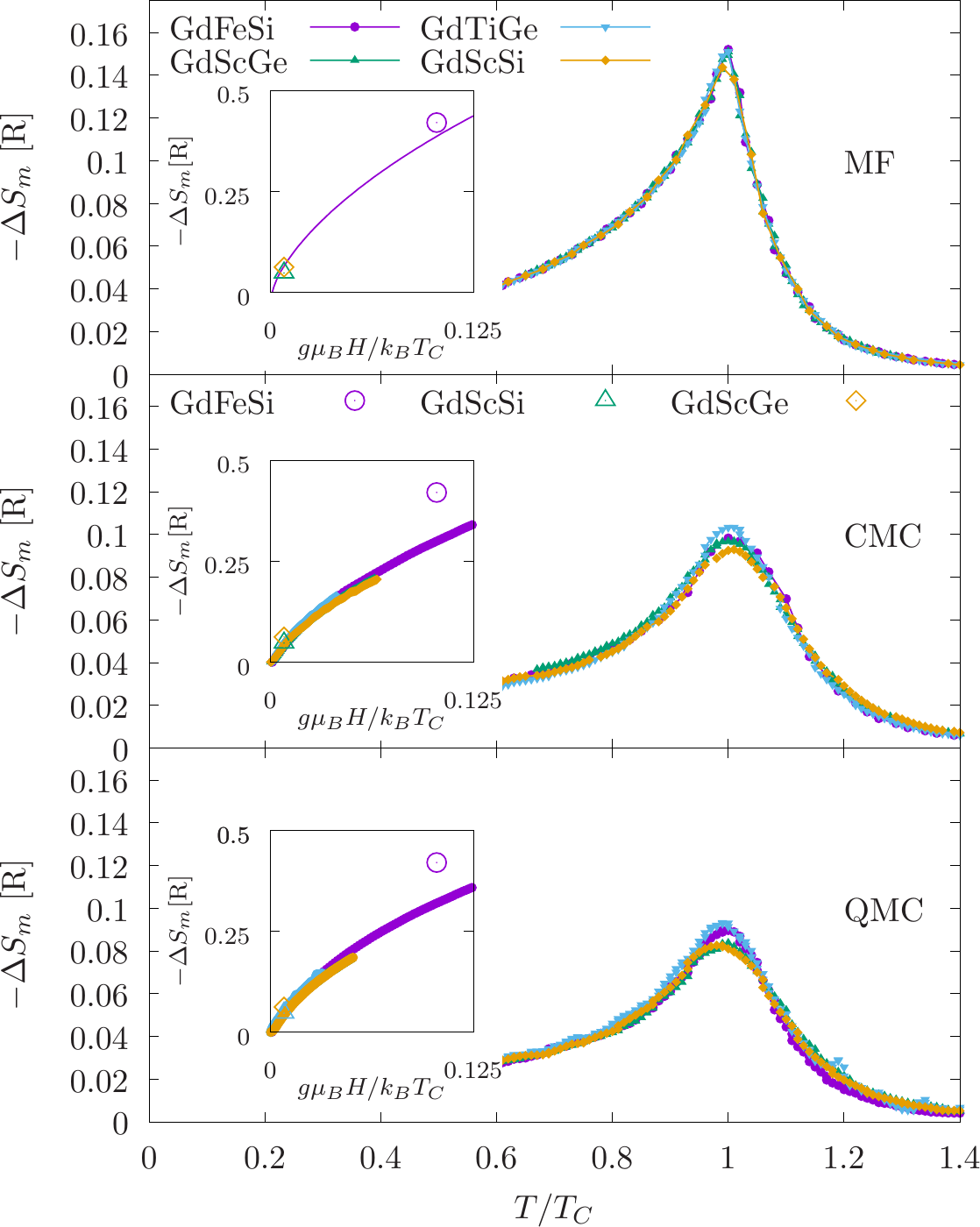}
    \end{center}
    \caption{(Color online) Magnetocaloric effect for different ferromagnetic compounds with both the temperature and the external magnetic field scaled by the corresponding $T_C$. The external magnetic field is $H=2\, \textrm{Tesla}$ for GdFeSi and $H=2\, \textrm{Tesla} (T_C/T_C^{GdFeSi})$ for the other compounds. The insets show the entropy change as a function of the rescaled field at $T_C$. The open symbols correspond to experimetal results with the magnetic field scaled by the experimetal transition temperature.}
    \label{fig:mecres}
\end{figure}

The scaling behavior is approximately followed in the CMC and the QMC calculations (see Fig. \ref{fig:mecres}). The maximum value of the entropy difference for a given external field is lower in CMC and QMC than in MF which is probably due to the lack of energy fluctuations at temperatures larger than $T_C$ in the MF approximation. 
The experimental entropy changes for GdFeSi, GdScGe and GdScSi are also shown in the insets of Figure \ref{fig:mecres}. At low (rescaled by $k_B T_C$) fields the experimental results are in very good agreement with the theory.
The large field result available for GdFeSi is larger than what is expected from the theory, which could be due to additional magnetic degrees of freedom not considered in our model (associated with the conduction band electrons).

\section{Conclusions}
We studied the magnetocaloric properties of Gd based RTX compounds having the CeScSi-type or CeFeSi-type crystal structures. Based on density functional theory calculations we obtained the ground state magnetic configuration and the exchange couplings of a simplified magnetic Hamiltonian. The lowest energy magnetic configurations obtained were in agreement with the available experimental data and the transition temperatures obtained consistent with the reported values. We found a weak dependence of the magnetic properties upon Si$\leftrightarrow$Ge replacement but a strong dependence of the inter-bilayer exchange coupling with the replacements Fe$\to$Co and Ti$\to$Sc that can even lead to a change of its sign and of the magnetic ground state configuration. 
The different behaviour observed in the T vs X replacements has to do with the fact that, in the first case, a change in the parity of the number of electrons in the 3d transition metal ocurrs while, in the second case, Si and Ge are isoelectronic.

A wide range of RTX compounds that share the CeFeSi-type crystal structure present the same qualitative change in the transition temperatures upon T replacement and X replacement. This is shown in Table \ref{tab:RTX}. Based on the R=Gd results we expect a very weak change in the magnetic exchange couplings in these compounds upon Si$\leftrightarrow$Ge replacement and a change in the sign of the inter bilayer coupling when Fe is replaced by Co.
\begin{table}
\begin{centering}
\begin{tabular}{l|cccccccccc}
\hline
\hline
	&	Ce&		Nd&	Sm&	 Gd&	Tb&	Dy&	Ho& 	Er&	Tm\\
\hline
RTiSi	&	-&		-&	-&	400&	286&	170&	95&	50&	20\\
RTiGe	&	-&		150&	260&	412&	270&	170&	115&	41&	15\\
RFeSi	&	-&	\cellcolor{Gray}25&	\cellcolor{Gray}40&	\cellcolor{Gray}118&	\cellcolor{Gray}125&	\cellcolor{Gray}110&	\cellcolor{Gray}29&	\cellcolor{Gray}22&	-\\
RCoSi	&	8.8&		7&	15&	175&	140&	-&	-&	-&	-\\
RCoGe	&	5&		8&	-&	-&	-&	-&	-&	-&	-\\
\hline
\hline
\end{tabular}
	\caption{Magnetic transition temperatures for compounds with the CeFeSi-type crystal structure. Shaded cells correspond to Curie temperatures.}
	\label{tab:RTX}
\end{centering}
\end{table}

We also studied the magnetocaloric properties of the R=Gd compounds and found a universal behavior of the magnetocaloric effect as a function of the temperature for the ferromagnetic compounds when the external magnetic field and the temperature are scaled by the transition temperature of each compound. This result, which is exact in the MF theory, and approximate in CMC and QMC sets a limit to the maximum MCE that can be expected in these compounds for a given $T_C$ and external magnetic field.

\section*{Acknowledgements}

We acknowledge financial support from PICT 2016-0204.
\bibliography{references}

\end{document}